# X-rays from the Power Sources of the Cepheus A Star-Forming Region

Steven H. Pravdo
Jet Propulsion Laboratory, California Institute of Technology
306-431, 4800 Oak Grove Drive, Pasadena, CA 91109; spravdo@jpl.nasa.gov

Yohko Tsuboi & Akiko Uzawa
Department of Physics, Faculty of Science and Engineering, Chuo University, Kasuga 1-13-27, Bunkyo-ku, Tokyo 112-8551, Japan; tsuboi@phys.chuo-u.ac.jp, akiko@phys.chuo-u.ac.jp

&

Yuichiro Ezoe, Ph.D.
Department of Physics, Tokyo Metropolitan University,
1-1 Minami-Osawa, Hachioji, Tokyo 192-0397, Japan; ezoe@phys.metro-u.ac.jp

## ABSTRACT

We report an observation of X-ray emission from the exciting region of Cepheus A with the *Chandra*/ACIS instrument. What had been an unresolved X-ray source comprising the putative power sources is now resolved into at least 3 point-like sources, each with similar X-ray properties and differing radio and submillimeter properties. The sources are HW9, HW3c, and a new source that is undetected at other wavelengths "h10." They each have inferred X-ray luminosities $\geq 10^{31}$ erg $s^{-1}$ with hard spectra, T $\geq$ $10^7$ K, and high low-energy absorption equivalent to tens to as much as a hundred magnitudes of visual absorption. The star usually assumed to be the most massive and energetic, HW2, is not detected with an upper limit about 7 times lower than the detections. The X-rays may arise via thermal bremsstrahlung in diffuse emission regions associated with a gyrosynchrotron source for the radio emission, or they could arise from powerful stellar winds. We also analyzed the *Spitzer*/IRAC mid-IR observation from this star-formation region and present the X-ray results and mid-IR classifications of the nearby stars. HH 168 is not as underluminous in X-rays as previously reported.

# 1. INTRODUCTION

Cepheus A is a well-known star formation region (Sargent 1977) with anisotropic mass outflow (Rodríguez et al. 1980) at a distance of ~700 pc (Johnson 1957, Moscadelli et al. 2009; throughout this paper we use d = 700 pc). It consists of two main H II regions, Cep A East and Cep A West (Hughes & Wouterloot 1982). The eastern region is believed to contain the exciting source or sources that may be identified with one or more of the ~14 compact radio sources (Hughes & Wouterloot 1984) contained therein. Polarization results point toward one compact radio source, HW2, as a main power source (Casement & McLean 1996). Many studies (Torrelles et al. 1993, Rodríguez et al. 1994, Hughes, Cohen, & Garrington 1995, Garay et al. 1996, Goetz et al. 1998, Gómez et al. 1999, Rodríguez et al. 2001) show that outflows, radio jets, and excitation phenomena are associated with HW2, believed to be a B0.5 protostar from which arises perhaps one half of the region's total luminosity, 2.5 x $10^4$ $L_\odot$ (Beichman, Becklin, and Wynn-Williams 1979, Evans et al. 1981). Patel et al. (2005) interpreted observations of submillimeter continuum emission and line emission as a massive disk of dust and molecular gas, respectively, around HW2 in support of this model (Jiménez-Serra et al. 2007, Torrelles et al. 2007). Brogan et al. (2007) found four compact submillimeter continuum sources and line emission in and around HW2 and interpreted this as several protostars rather than a single massive disk (Comito et al. 2007). HW2 and the other non-varying radio sources are joined by two or more variable compact radio objects (Hughes 1991), all presumably within the confines of this same region.

The first X-ray observation with moderate spatial resolution (≤5") revealed an unresolved source of X-rays centered near HW2 with a luminosity of $L_X$ ~ 1.6 x $10^{31}$ ergs $s^{-1}$ and a temperature T ~ $10^8$ K (Pravdo & Tsuboi 2005, hereafter PT05). PT05 concluded that X-rays came from more than one source over a region several arcseconds in extent. The X-ray spectrum and luminosity is comparable to that from flaring low-mass young stellar objects (YSOs). In this paper we present the results of a Chandra/ACIS observation of the region that features higher spatial resolution (≤1") and comparable sensitivity.

The Cepheus A West region also contains compact radio sources (Hughes & Moriarty-Shieven 1990) and the Herbig Haro object HH 168 (GGD 37, Gyulbudaghian et. al. 1978). The latter shows the characteristics of other energetic HH optical condensations with high excitation and linewidths (Hartigan, et al. 1986) and high velocities and proper motions (Lenzen 1988). Hartigan & Lada (1985) identified several optical knots within HH 168 that were resolved and further analyzed with Hubble Space Telescope (HST) spectroscopy (Hartigan, Morse, & Bally 2000). PT05 found soft X-ray emission (<1 keV) associated with HH 168.

# 2. OBSERVATIONS AND ANALYSIS

### *2.1 Chandra/ACIS*

We performed a 78.0-ks observation with the *Chandra* X-ray Observatory (Weisskopf et al. 2002) on April 8-9, 2008 with the ACIS-I imaging array (CCDs I0-I3) in the standard mode. Data were reduced using the latest version of the CIAO analysis software.[1] In this paper we discuss the measurements from CCD I3 only, which contains the exciting region of Cepheus A East and HH 168.

We performed positional, timing, and spectral analyses of these data. We used a wavelet-based source detection program (Freeman et al. 2002) to locate X-ray sources on CCD I3 in a soft band, 0.4-2 keV, and in a hard band, 2-10 keV. We performed encircled energy analysis by examining the spatial distribution of detected sources around their centroids. We constructed light curves to check for source variability. The degree of variability was determined using the Gregory-Loredo (G-L) algorithm (Gregory & Loredo 1992).

Finally, we used XSPEC to measure the spectra of the brightest detected sources. Source counts were extracted from regions around sources that matched the local point-spread function (PSF), and backgrounds were taken from nearby source-free regions. The instrumental response matrices are deemed to be reliable > 0.4 keV, so we only examined spectra in the range 0.4-10 keV. Above 10 keV there were no significant counts. The spectral models included the functions "WABS" for interstellar absorption and "MEKA-L" for a thermal spectrum (Mewe, Gronenschild, & van den Oord 1985). All of the spectral models described below had acceptable fits to the data with reduced $X^2$ ~1. We calculated the 1-σ limits for spectral parameters by first allowing all the parameters to vary simultaneously to find the best-fit values. Then a parameter of interest was fixed and we stepped through different values of that parameter, allowing all the other parameters to vary. The value of the interesting parameter when chi squared increased by 3.5 is the 1-σ limit for a fit with 3 free parameters (Lampton, Margon, & Bowyer 1976). We obtained the same results using the C-statistic.

We checked the astrometric accuracy of the X-ray data by comparing the positional output of the source detection program with the positions of unambiguously identified stellar counterparts found in the 2MASS (Skrutskie et al. 2006) catalog. For 18 X-ray sources that had positional coincidence of less than 0.5" with 2MASS objects, the average and standard deviations of the RA and Decl. offsets were -0.08± 0.19" and -0.01 ± 0.19", respectively.

Tables 1 and 2 list the X-ray sources detected on CCD I3. In Table 1 we list the 39 X-ray sources that are known at other wavelengths or previously known in X-rays, and their counterparts. The "source id" (Column 1) begins with "h" if the source is seen in the hard or both bands and begins with "s" if the source is only seen in the soft band. Columns 2 and 3 give the X-ray source position. Columns 4 and 5 give the hard and soft count rate, respectively, and column 6 gives the X-ray flux assuming a distance of 700 pc. The fluxes are determined from spectral fits for the brightest sources (see Table 3) and scaled for the other sources. We also show in column 7 the X-ray flux determined from PT05, scaled to the same source distance. All positional associations with USNO B1 (Monet et al. 2003, column 8), 2MASS (column 9), and newly identified IRAC sources (columns 10-13) are within 1" except for those noted. In Table 2 we list 13 "blankfield" X-ray sources, i.e. with no counterparts in other bands. Fig. 1 shows the location of the sources in the *Chandra/ACIS* image. The properties of HH 168 (see below) are in Table 3 only.

The possibility exists that extragalactic X-ray sources penetrate this star formation region and are detected in our observation. Feigelson et al. (2002) estimate that in a similar observation ≤ 2% of the sources are extragalactic contaminants. This implies that ≤ 1 of our sources is from the extragalactic background.

[1] http://cxc.harvard.edu/ciao/index.html

## 2.2 *Spitzer*/IRAC

We also make use of *Spitzer* Space Telescope (Werner et al. 2004) archived data from the Infrared Array Camera (IRAC, Fazio et al. 2004). We examined the data from the Multiband Imaging Photometer for *Spitzer* (MIPS, Rieke et al. 2004) but saturation rendered the data unusable. The observation was performed on July 23, 2004. We created deconvolved images and outlier masks with the *Spitzer* Science Center (SSC) Mosaicking and Point-source Extraction (MOPEX) software package (Makovoz & Marleau 2005). We applied the outlier masks, along with Data Collection Event and Permanent Bad Pixel masks from the SSC pipeline. The sources are oversampled with 0.25 arcsec pixels, to match the deconvolved images.

We created enhanced-resolution images in the four bands - 3.6 , 4.5, 5.8, and 8.0 μm - using the HiRES software package developed at JPL (Backus et al. 2005), and IRAC point-response-functions (PRFs) distributed through an SSC-contributed software package, "Iracworks"[2]. Resolution is enhanced by a factor of ~3 with HiRES software (Velusamy, Langer, & Marsh 2007) and the algorithm properly conserves flux (Backus et al. 2005). Although this results in ringing around bright point sources with extended background emission, it does concentrate the flux in the Airy rings for point sources, and better defines the shape of extended sources, and thus is best for deblending crowded sources, or distinguishing the flux of a point source from the background around it.

We performed source extraction on the images, using the SExtractor software (Bertin & Amouts 1996). This returns an integrated flux for each source in each image. We calculated both an aperture flux within a fixed circular aperture tuned to the known point-spread function (which we report herein) and an isophotal flux (for a check). We also examined the elongations of the images since a point-like source would have an elongation close to 1.

To verify our IRAC astrometry we compared these objects with coincident X-ray sources within 0.5". For 16 sources in the IRAC 3.6-μm band the offsets were 0.20 ± 0.18" in RA and 0.10 ± 0.16" in Decl. Comparing X-ray sources with IRAC 8.0 μm sources yields offsets of 0.08 ± 0.21" in RA and 0.06 ± 0.17" for 14 sources.

Figures 2a-d shows the Cepheus A region in the mid-IR from the four IRAC bands with the X-ray source positions overlaid. The emission is concentrated in two clusters: one in the southeast contains the power sources and the other in the northwest contains a number of bright X-ray sources (see below). Table 1 also gives the fluxes from the fixed circular aperture in mJy from the sources with IRAC counterparts. In a couple of noted cases we use the isophotal flux when the aperture flux is not well-determined, or we estimate the flux when the automatic processing fails due to source confusion. For 29 sources we constructed spectral energy distributions (SEDs) and divide them into YSO classes by computing their SED indices according to the method of Greene et al. (1994). Following Prisinzano et al. (2008) we classify sources with rising IRAC SEDs in all bands as Class 0-I since we do not have spectral data above 8 μm to distinguish Class 0

[2] http://ssc.spitzer.caltech.edu/archanaly/contributed/browse.html

from Class I. These indices and classes are also listed in Table 1 (columns 14 and 15, respectively). The Cepheus A East core in the mid-IR is shown in an expanded view in Fig. 3a-d.

### *2.3 The Cepheus A East Core*

Figures 4a and 4b show the X-ray image of the exciting region of Cepheus A East. The central region that was previously reported to contain the unresolved X-ray source "HWX" (PT05) is now seen to consist of at least 3 X-ray sources, shown with the larger green and cyan circles labeled h9-h10-h11 in Fig 4a. Also shown in the figures (smaller magenta circles) are the positions of the compact "HW" radio sources (Garay et al. 1996). Fig. 4b is an expanded view of the central region and shows higher resolution radio observations of Torrelles et al. (1998)—0.2"-radius green circles--as well as the submillimeter sources discovered by Brogan et al (2007)—0.1"-radius cyan circles. At the ≤1" spatial resolution of the X-ray measurements there is a coincidence between two X-ray sources detected with the wavelet algorithm and two HW sources, viz., HW9 with h9 and HW3c with h11. In addition the submillimeter array source SMA HW3c is also coincident with h11. The third X-ray source, h10 has no corresponding source at other wavelengths. None of the other HW or submillimeter sources, including HW2, have a corresponding X-ray source detected with the standard analysis. However, there are excesses of X-ray emission that may be associated with HW8 and HW3b.

We analyzed the spectra of the three X-ray sources associated with this central region with results shown in Table 3. The net X-ray counts are about the same in each source, between 51 and 62. The spectral parameters are not well-determined but all three fit well to a high-temperature MEKAL thermal model that is highly absorbed at low energies. To get better statistics we formed a total X-ray spectrum as a guide, with the assumption that the underlying source spectra are similar. This spectrum (Table 4, Fig. 5) fits the absorbed MEKAL model well, $X^2$ = 31 for 35 degrees of freedom (d.o.f.), with T = 8.7 (+8.7,-4.3) x $10^7$ K with $N_H$ = 2.5 (+1.1,-0.6) x $10^{23}$ H $cm^{-2}$, which corresponds to 113 (+41,-32) visual magnitudes of absorption. This is consistent with the temperature found by PT05 for the composite source in 2003 and ~3 times more absorbed than their report of 36 (+14,-9) visual magnitudes of absorption. There is also evidence for iron Kα line emission near 6.7 keV consistent with the possibility that the spectrum is that of plasma in thermal equilibrium with cosmic abundances.

The total observed intensity of h9-h10-h11 is 9.3-11 x $10^{-14}$ ergs $cm^{-2}$ $s^{-1}$, depending upon whether we use the individual spectra (Table 3) or the total spectrum (Table 4). This is comparable to the HWX intensity of 13.1 x $10^{-14}$ ergs $cm^{-2}$ $s^{-1}$ from PT05. However because of the higher absorption measured herein, the unabsorbed luminosity is 2.2 (total spectrum) or 7.9 (individual spectra) x $10^{31}$ ergs $s^{-1}$ compared with 1.6 x $10^{31}$ ergs $s^{-1}$ (PT05), or as many as 5 times higher than the luminosity of HWX, five years earlier. Note that in this paper we use the slightly smaller distance estimate of 700 rather than 725 pc, resulting in an overall decrease in the luminosities of ~7% from the earlier estimates. This possible increase in the luminosity from 2003 to 2008 is uncertain because the large effects of low-energy absorption result in uncertainties in the estimates of the unabsorbed fluxes.

Some of the sources appear to be variable by a factor of ≤ 5, with HW9 "definitely variable" and HW3c "considered variable" by the G-L algorithm. In contrast h10 is considered to show "no hint of variability." Figures 6a-c show the light curves of HW9, HW3c, and h10, respectively.

### *2.4 Other X-ray Sources*

The other X-ray sources that surround the central core appear to be typical of those in star formation regions. Those identified by class based upon their IRAC fluxes include 3 Class 0-I, 13 Class II, 11 Class III, and 2 flat-spectrum stars, typical of star formation regions (e.g. Broos et al 2007, Getman et al. 2007, Pravdo et al. 2009, and many references therein). Three bright X-ray sources are located in the northwest cluster of enhanced mid-IR emission shown in Fig. 2a-d. This cluster of emission generally brightens from 3 to 8 microns and is larger, although not brighter than the power source cluster to the southeast at 8 microns (Fig. 2d). The brightest X-ray source, as it was in PT05, is HL 8 or h38 (Hartigan & Lada 1985), now identified as a Class II object from IRAC spectrophotometry. The 2008 X-ray spectrum (Table 3, Fig. 7) is similar to what it was in 2003 and the X-ray luminosity is within a factor ~2. The lights curves of HL 8 and the nearby bright X-ray source Class II object HL 9 (h36), are shown in Fig. 6d-e. Both are considered "definitely not variable" by the G-L algorithm. The final light curve shown in Fig. 6f is for h37, another bright X-ray source that shows a flare and is "definitely variable." This is also likely to be a Class II object although source confusion (Fig. 2) did not allow us to unambiguously extract its mid-IR spectrum.

There are three newly identified Class 0-I sources: h19, h12, and h7, all on the northern periphery of the southeast power-source nebula. One of these, h7, was detected before in X-rays by PT05, and another, h19, is associated with the radio source C1 (Hughes & Moriarty-Shieven 1990). All three Class 0-I stars have rising SED in all IRAC bands and are not detected in the visible (or 2MASS) and are thus Class 0-Ia as defined by Prisinzano et al. (2008). They have low net counts and we can only conclude that individually they have hard spectra (Table 3) similar to Class I objects in other star-formation regions (e.g. Ozawa et al. 1999). We formed and analyzed a total spectrum for these sources and display the result in Table 4. We also report in Table 4 the fits to total spectra for Class II sources (other than those listed in Table 3), for the Class III sources with hard components, and separately for the blankfield sources with hard and soft components.

### *2.5 HH 168*

This observation was not optimized to view the soft X-ray emission from HH 168 because CCD I3 is not as sensitive <2 keV as the back-side illuminated CCD 7. Nevertheless Figure 8 shows soft X-ray emission (red) near the locations of the PT05 localizations of HH 168 (white squares). The image is smoothed over 10 pixels. The emission seen with Chandra/ACIS appears to be diffuse. X-ray contours are shown in Fig. 9 superposed on the Hα image (Hartigan et al. 2000). The X-ray contours look similar in many respects to those shown in Fig. 4 of PT05, despite the differing detectors and resolutions. Both cover much of the extent of the Hα emission although the contours in the earlier image appeared to extend farther west. Both have local peaks in X-ray emission near the location of the bright HH objects HW and E. The previous localizations of the X-ray emission fall within the new contours but slightly displaced. The integrated X-ray emission is larger than reported by PT05 with an unabsorbed intensity of 3.0 x $10^{30}$ erg $s^{-1}$. A re-examination of the PT05 data shows that the previous flux was incorrect and is in fact consistent with the current value. Fig. 10 shows the simultaneous fit of the Chandra and XMM data. The model has a good fit (reduced $X^2$ = 1.0 for 68 d.o.f.) The X-ray spectrum (Fig. 10) is relatively soft with a temperature of 6.5 (+0.8,-0.6) x $10^6$ K and low energy absorption of 5.1 (+1.7,-3.3) x $10^{21}$ H $cm^{-2}$ (Table 3). The facts that the flux and spectrum did not change, the relatively low X-ray temperature, and the source extension imply that the HH object and the many sub-objects are the major sources of the X-rays, rather than the extant T Tauri stars (Hartigan et al. 2000).

## 3. X-RAYS FROM THE CEPHEUS A POWER SOURCES

The X-ray emission from the central region of Cepheus A East is now resolved into at least three point-like sources. Despite the apparent similarity in their X-ray properties (Tables 1, 3, and section 2.3), albeit based upon only 50-60 counts each, each of these has very different associations at other wavelengths. In particular, one is a variable radio source-HW9; one is a steady-state radio source-HW3c; and one is not detected in the radio-h10. Upper limits to the X-ray emission from other radio sources including HW2, considered to be the most massive protostar in the complex, are ~7 times lower than the detections. All three X-ray spectra are heavily absorbed at low energies implying high column densities of intervening materials that correspond to minimum of 23 magnitudes of visual absorption. Their composite spectrum most resembles that of the Class 0-I sources (Table 4). Class II and Class III sources exhibit not only considerably less low-energy absorption, but also lower temperatures, as already reported in other star-forning regions (e.g. Imanishi et al. 2001). While the former is probably indicative of their environment and the line of sight, the latter could be an intrinsic source difference. The possibility exists that one or more of these are Class 0 protostars. HW3c in particular seems to share many of the X-ray and submillimeter properties of other putative members of this class (Tsuboi et al. 2001).

What can be said about the masses of these three stars? The radio variability and now the X-ray variability of HW9 point toward its being a low-mass PMS star. HW3c has been considered a higher mass star because of its bright, steady state radio and sub-millimeter emission, but its apparent X-ray variability may either point toward a lower mass or may tell us that high mass stars have short-term X-ray variability. The nature of h10, seen only in X-rays, is unknown.

There is no evidence for the hard X-ray ridge reported by PT 05 along the HW7a-d chain of radio sources. However, the X-ray source h6, associated with a "flat" spectrum mid-IR object, is near HW7a. It is also located at the end of the HW7b-d string. These are usually believed to be associated with power sources closer to the radio center but the proximity of this new object, perhaps a low-mass PMS star, raises a question as to their origin. The X-rays are highly absorbed and the source has a large inferred luminosity (Table 3).

### *3.1 HW9*

HW 9 is a highly variable compact radio source that made its appearance above the detectability limit several years after sensitive radio-monitoring of the Cepheus A region began. Flaring was then measured between observations to be at least of factor of ~6 within 57 days (Hughes 1991). Hughes (1991) suggested that the source is a protostar or pre-main sequence (PMS) B3 star. He interpreted the radio flux and variability as arising from gyrosynchrotron emission in the surrounding nebula from a region, e.g., ~1 AU in size with T ~ $10^8$ K, B ~200 G, and electron or ion density, $n \sim 10^6$ $cm^{-3}$. The variability arises because of magnetic field and/or temperature changes. In such a model we can compute the emissivity due to thermal bremsstrahlung; viz. ~1.5 x $10^{-14}$ $n^2$ $(T/10^8)^{½}$ erg $cm^{-3}$ $s^{-1}$, which results in $L_X \sim 2 \times 10^{29}$ erg $s^{-1}$. One could attain the inferred $L_X = 2.4 \times 10^{31}$ erg $s^{-1}$ by increasing the assumed size by 5 or density by 10, or some combination of the two, neither of which violates the constraints. The internal column density of this region would be $N_H \leq 10^{20}$ $cm^{-2}$ consistent with the observed X-ray absorption being external to the source. This is a natural way to link and explain the radio and X-ray emission and would predict the presence of thermal line emission such as the possible Fe Kα line in the spectrum (Fig. 5). A possible downside to this model is the variability time scale. There is no useful lower limit to the radio variability but it may be well below the observed ~57 days. Possible variability in the X-ray emission has a time scale of hours (Fig. 6a) but requires confirmation.

If short-term variability rules out a large emission region for the X-rays, then we must look closer to the star to find their origin. Other X-ray producing mechanisms such as from a high velocity stellar wind or an unseen companion are still viable. Hughes, Cohen, & Garrington (1995) show that HW9 is at the center of disrupted high-velocity gas and conclude that may be evidence for such a wind. The stellar wind hypothesis implies a shock velocity, $v_s \geq (16kT/3\mu m_p)^{½} = 510$ km $s^{-1}$, where $T$ is the observed X-ray lower temperature limit, $m_p$ is the proton mass, and $\mu$ is the mean molecular mass in the wind (e.g. Pravdo et al. 2001). This and higher velocities are attainable in B-star stellar winds (Cassinelli et al. 1994). Alternatively, the X-ray luminosity of HW9 is large but not unprecedented for identification as a low-mass PMS star (Garay et al. 1996). A scaled up version of a solar flare model consisting of both thermal plasma and nonthermal electrons (Bastian et al 2007) may be applicable if coordinated X-ray and radio observations reveal correlated short term variability. However, such a correlation has not been observed in at least one other star formation region (Osten & Wolk 2009). The ratio of X-ray to radio flux $L_X/L_R = 0.2\text{-}1.4 \times 10^{14}$ Hz is on the low end of the "quiescent" emission ratio for several classes of active stars including weak-lined T Tauris and solar flares (Güdel 2002), although applicability of this comparison is called into question by the radio variability of HW9.

### *3.2 HW3c*

HW3c is a persistent compact radio source (Hughes & Wouterlout 1984) that has been observed to be moderately variable, ~20% (Hughes 2001). It is one of the two compact radio sources that is detected as a strong submillimeter source, the other being HW2. It is now clear that HW3c is a strong hard-X-ray source whereas HW2 is not. Hughes (2001) proposes that there are two components in the radio emission: synchrotron radiation between 1.4-4.9 GHz from ionized jets and gyrosynchrotron emission between 4.9-43 GHz similar to the model for HW9 above. The jets are likely to be associated with a powerful outflow that terminates in bright HH 168 emission to the west and a fainter $H_2$ bow to the east (Cunningham, Moeckel, & Bally 2009). Where might the X-rays arise? X-ray emission from jets is now well known (Pravdo et al. 2001, Favata et al. 2002, Pravdo, Tsuboi, & Maeda 2004). However jet X-rays have temperatures ~$10^6$ K indicative of shocked material at velocity ≥200 km $s^{-1}$. This is very different than the X-rays observed herein. Rather the X-ray temperature is reminiscent of the temperature associated with the putative gyrosynchrotron region or of the processes associated with the underlying protostar itself, as discussed above for HW9.

The case that HW3c is a massive star (e.g. Hughes 1988) is more secure than for HW9 due the former's submillimeter emission and associated outflows. Nevertheless their X-ray emission is similar at this statistical level. In this case the radio emission is reasonably steady state so the low value of the ratio $L_X/L_R$ = 3.2 x $10^{13}$ Hz may be a clearer indication that HW3c is not among the classes of lower-mass stars in Güdel's (2002, fig. 6) general trend.

### *3.3 h10*

This source is similar in X-ray intensity and spectrum to its two neighbors, HW9 and HW3c. It inferred luminosity is considerably lower. However, this source differs from the others because it has not been detected in either radio or submillimeter. It may have been a component of the HWX composite source in PT05, as indeed, all of these three may have been.

## 4. COMPARATIVE TAXONOMY

The variety of source types in Cepheus A East challenges our ability to categorize them. The compact radio objects and now the X-ray and mid-IR sources overlap in a still unsolved way. For comparison we consider a similarly diverse star-formation region containing another "buried" power source, GGD 27. This region is ~2.5 times further away but has been studied with comparable broadband detail. Leaving aside HH objects and other outflow sources, the only compact radio sources detected in the X-rays were MRR 12, 32, and 14 (Marti, Rodríguez, & Reipurth 1993). The first two were shown to be Class III PMS stars (Pravdo et al. 2009). In Cepheus A East none of the classified mid-IR sources is a radio source. Could the difference for MRR 12 and 32 be a strong magnetic field, such as was invoked to explain the radio emission from HW9?

The last of the radio sources in GGD 27 is the putative power source, GGD 27-ILL = GGD 27-X. The X-ray emission is similar to that associated with the power sources herein: hard with large low-energy absorption and an inferred luminosity ≥ $10^{31}$ erg $s^{-1}$. This object is thought to be a massive B0.5 protostar with an outflow. Pravdo et al. (2009, Table 3) list the small number of other similar X-ray objects—four including GGD 27-X. The object herein that most fits into this category is HW2, also considered to be a B0.5 PMS star with a massive outflow and a bolometric luminosity > $10^4$ $L_\odot$ (Rodríguez et al. 1994 and references therein). But HW2 is not an X-ray source. This is rather surprising when one considers that X-rays are detected from the three nearby stars described above. If an underlying X-ray source in HW2 is similar to the other power sources then there must be dramatic spatial gradient in the X-ray-absorbing column density between HW3c and HW2. If the column density scales with the mass of gas inferred from the submillimeter observations (Brogan et al. 2007), then a factor of ~6 can be explained. A more sensitive observation at X-ray energies > 5 keV would test this hypothesis. Future observations may reveal, however, that there is significant large amplitude variability in these sources, and that we have caught HW2 in a quiescent phase.

## 5. HH OBJECTS IN X-RAYS AND RADIO

Figure 11 shows a summary plot of the radio and unabsorbed X-ray emission from HH objects with the corrected X-ray luminosity for HH 168. The HH 168 radio luminosity is taken only from that of HH object E (Hughes & Moriarty-Schieven 1990) and objects W1-W3 (Garay et al. 1996), all of which are included within the boundaries of the diffuse X-ray emission that we observed. The best-fit linear relationship shows $L_R \sim L_X^{0.7}$, which is flatter than the index ~1.24 that characterizes a large range of active stars and flares described by Güdel (2002). In the case of HH objects both emission regimes likely arise from free-free emission in different, but related, temperature and density regions of shocks formed by the collisions of outflowing jets and the ambient interstellar medium. For the radio the shock temperatures are $\sim 10^4$ K whereas the X-ray temperatures reach $\sim 10^6$ K, derived from the velocities observed via proper motions. A possible explanation of the radio-X-ray relationship shown in Figure 11 is that the X-rays arise from non-radiative bow shocks (e.g. Raga et al. 2002) and thus depend upon the square of the density, whereas the radio emission depends only linearly on the density (e.g. Curiel et al. 1993). Hence the X-ray emission grows at a faster pace as the densities and luminosities grow.

## 6. CONCLUSIONS

This first high-resolution and moderate sensitivity X-ray observation of Cepheus A East resolves the X-ray emission from the region of the power sources into at least three diverse luminous X-ray objects. The sources run the gamut of X-ray properties from variable to steady-state to undetected. Surprisingly, the most energetic source at other wavelengths is not detected in X-rays. The X-ray emission is associated with PMS and protostars. Further observations of short-term variability may distinguish between models of extended emission or models closer into the stars such as strong stellar winds.

**ACKNOWLEDGEMENTS**
The research described in this paper was performed in part by the Jet Propulsion Laboratory, California Institute of Technology, under contract with the National Aeronautics and Space Administration. We thank T. Thompson for assistance with the *Spitzer* data reduction, and G. Garmire and P. Broos for assistance with the *Chandra* data reduction. This research has made use of the NASA/ IPAC Infrared Science Archive including the *Spitzer* archive, which is operated by the Jet Propulsion Laboratory, California Institute of Technology, under contract with the National Aeronautics and Space Administration. This research has made use of the SIMBAD database, operated at CDS, Strasbourg, France. This publication makes use of data products from the Two Micron All Sky Survey, which is a joint project of the University of Massachusetts and the Infrared Processing and Analysis Center/California Institute of Technology, funded by the National Aeronautics and Space Administration and the National Science Foundation. This research was supported by NASA contract NAS 8-01128. Government sponsorship acknowledged. Y.T. acknowledges support from the Grants-in-Aid for Scientific Research (number 20540237) by the Ministry of Education, Culture, Sports, Science and Technology. Y.T. also acknowledges T. Maeda, Y. Sugawara and H. Kobayashi for help with the data analysis. 

Table 1. X-ray Source Properties and Identifications

| Source ID | RA (2000) | Decl. (2000) | 2-10 keV cts $ks^{-1}$ | 0.4-2 keV cts $ks^{-1}$ | 2008[e] $L_X$ $10^{30}$ erg $s^{-1}$ | 2003[e] $L_X$ $10^{30}$ erg $s^{-1}$ | USNO B1 | 2MASS | IRAC1 mJy | IRAC2 mJy | IRAC3 mJy | IRAC4 mJy | Index | Class | Notes |
|---|---|---|---|---|---|---|---|---|---|---|---|---|---|---|---|
| h41 | 343.97544 | 62.05873 | 0.38+/-0.07 | 0.3+/-0.06 | 0.82 | - | 1520-0380814 | 22555409+6203313 | 3.8 | 3.6 | 2.2 | 1.3 | -2.4 | III | - |
| h40 | 343.98805 | 62.04145 | 0.13+/-0.04 | 0.08+/-0.03 | 0.25 | - | - | 22555714+6202289 | 7 | 5.9 | 3.3 | - | -2.6 | III | - |
| s23 | 343.99960 | 62.04273 | - | 0.37+/-0.07 | 0.15 | - | 1520-0380828 | 22555991+6202336 | 0.3 | - | - | - | - | - | - |
| h38 | 344.00761 | 62.05457 | 5.63+/-0.27 | 8.54+/-0.34 | 16 | 7.6 | 1520-0380840 | 22560182+6203165 | 54 | 64.5 | 72.8 | 88.4 | -0.4 | II | XMMPT J225601.6+620316, HL 8 |
| h37 | 344.01207 | 62.04930 | 3.83+/-0.23 | 0.61+/-0.09 | 10 | - | - | 22560297+6202574 | - | - | 22.6 | - | - | - | - |
| h36 | 344.01873 | 62.05084 | 2.14+/-0.17 | 2.62+/-0.19 | 6.4 | 5.2 | 1520-0380855 | 22560450+6203030 | 32.8 | 23 | 49.7 | 41.4 | -0.4 | II | XMMPT J225604.3+620303, HL9 |
| h35 | 344.02303 | 62.05190 | 1.23+/-0.13 | 0.52+/-0.08 | 2.2 | - | 1520-0380860 | 22560553+6203068 | 20.2 | 17.8 | 17 | 21.8 | -0.9 | II | - |
| h34 | 344.02358 | 62.04486 | 0.26+/-0.06 | 0.38+/-0.07 | 1.2 | - | - | 22560563+6202415 | 7.8 | 4.9 | - | - | -3.0 | III | - |
| h33 | 344.02831 | 62.07599 | 0.1+/-0.04 | 0.05+/-0.03 | 0.19 | 3.3 | - | 22560687+6204336 | - | - | - | - | - | - | XMMPT J225606.7+620433 |
| h32 | 344.03052 | 62.04134 | 2.98+/-0.2 | 0.7+/-0.1 | 12 | 2.8 | - | 22560733+6202287 | 11.9 | 19.5 | 16.6 | 11.5 | -1.2 | II | XMMPT J225607.1+620230 |
| h31 | 344.03178 | 62.05876 | 1.07+/-0.12 | 0.49+/-0.08 | 2.0 | 0.86 | 1520-0380873 | 22560761+6203307 | 41.9 | 42.4 | - | - | -0.9 | II | XMMPT J225607.4+620330 |
| h29 | 344.03777 | 62.05954 | 0.08+/-0.03 | 0.1+/-0.04 | 0.20 | - | - | 22560911+6203346 | 1.7 | 1.3 | - | - | -2.2 | III | - |
| s29 | 344.04159 | 62.06339 | - | 0.1+/-0.04 | 0.10 | - | - | 22561002+6203481 | 0.9 | 0.7 | - | - | -2.3 | III | - |
| h28 | 344.04707 | 62.03634 | 0.29+/-0.06 | 0.21+/-0.05 | 0.60 | 0.95 | - | 22561127+6202111 | 22.5 | 24.9 | 24.2 | 25.2 | -0.9 | II | XMMPT J225611.5+620213, HL 12, IRS4 |
| h27 | 344.04793 | 62.02807 | 0.21+/-0.06 | - | 0.30 | - | - | 22561153+6201408 | 1.5 | 1.6 | 1.0 | - | -1.9 | III | - |
| s10 | 344.04973 | 62.00980 | - | 0.11+/-0.04 | 0.10 | - | - | 22561192+6200357 | 1.6 | 1.1 | 0.7 | - | -2.7 | III | - |
| h25 | 344.05009 | 62.03842 | 0.12+/-0.04 | 0.11+/-0.04 | 0.28 | - | - | 22561201+6202187 | 0.8 | - | - | - | - | - | - |
| h24 | 344.05135 | 62.01340 | 0.24+/-0.06 | 0.35+/-0.07 | 0.68 | - | - | 22561239+6200482 | 7.4 | 3.8 | 3.9 | 1.7 | -2.7 | III | KCW94 rad |

| Source ID | RA (2000) | Decl. (2000) | 2-10 keV cts ks$^{-1}$ | 0.4-2 keV cts ks$^{-1}$ | 2008 $L_X$ $10^{30}$ erg s$^{-1}$ | 2003 $L_X$ $10^{30}$ erg s$^{-1}$ | USNO B1 | 2MASS | IRAC1 mJy | IRAC2 mJy | IRAC3 mJy | IRAC4 mJy | Index | Class | Notes |
|---|---|---|---|---|---|---|---|---|---|---|---|---|---|---|---|
| h42 | 344.05360 | 62.04513 | 0.17+/-0.05 | 0.15+/-0.05 | 0.39 | 1.0 | - | 22561302+6202425 | 4.8 | 3.5 | 2.3 | 3.3 | -1.5 | II | XMMPT J225612.7+620240 |
| h23 | 344.05417 | 62.02012 | 0.18+/-0.05 | - | 0.26 | - | - | 22561302+6201123 | 12 | 20.8 | 23.5 | 18.1 | -0.5 | II | - |
| h22 | 344.05443 | 62.01635 | 0.18+/-0.05 | - | 0.26 | 5.6 | - | 22561309+6200588 | 6.6 | 9 | 11.8 | 8.4 | -0.7 | II | XMMPT J225613.2+620059 |
| h20 | 344.05548 | 62.03041 | 1.71+/-0.15 | 0.47+/-0.08 | 9.4 | 8.0 | - | 22561333+6201494 | 6.8 | 3.9 | - | - | -3.5 | III | XMMPT J225613.4+620149 |
| h19 | 344.05822 | 62.03830 | 0.18+/-0.05 | - | 0.48 | - | - | - | 1.7 | 7.2[a] | 102.5 | 277 | 5.6 | 0-I | C1 (Hughes & Moriarty-Shieven 1990) |
| h18 | 344.05890 | 62.03485 | 0.2+/-0.05 | - | 0.28 | - | - | - | 0.5[b] | - | - | - | - | - | - |
| h16 | 344.06201 | 62.02339 | 0.76+/-0.1 | 0.88+/-0.11 | 1.90 | 1.6 | 1520-0380893 | 22561490+6201241 | 33.7[a] | 29.4 | 28.4 | 11.7 | -2.3 | III | XMMPT J225614.6+620125, LHS84 IRS5 EQ 225416+6202024-HL 20 |
| h15 | 344.06781 | 62.03846 | 0.68+/-0.1 | 0.26+/-0.06 | 1.21 | - | - | 22561627+6202185 | 37.6 | 47.2 | 45.4 | 38.1 | -1.0 | II | - |
| h13 | 344.07382 | 62.03992 | 0.2+/-0.05 | 0.23+/-0.06 | 0.50 | - | 1520-0380897 | 22561768+6202235 | 1.2[b] | 21.9[c] | 7.0[c] | 2.2 | -1.0 | II | HL 29 |
| h12 | 344.07422 | 62.04064 | 0.21+/-0.06 | - | 0.88 | - | - | - | 0.4 | 5.5 | 7.1 | 12.1 | 2.8 | 0-I | |
| h11 | 344.07458 | 62.02955 | 0.84+/-0.11 | - | 49 | 16[d] | - | - | - | - | - | - | - | - | HW3c-SMA, XMMPT J225618.4+620147 (HWX) |
| h10 | 344.07550 | 62.02873 | 0.68+/-0.1 | - | 6.2 | 16[d] | | | - | - | - | - | - | - | XMMPT J225618.4+620147 (HWX) |
| h9 | 344.07766 | 62.02991 | 0.73+/-0.1 | - | 24 | 16[d] | - | - | - | - | - | - | - | - | XMMPT J225618.4+620147 (HWX), HW9 |
| h7 | 344.08159 | 62.03981 | 0.71+/-0.1 | - | 13 | 1.1 | - | - | - | 0.4 | 10.3 | 13.8 | 4.9 | 0-I | XMMPT J225619.4+620223 |
| h5 | 344.08400 | 62.02396 | 0.36+/-0.07 | - | 0.51 | - | - | - | 12.9 | 22.8 | 25.1 | 20 | -0.5 | II | - |
| h6 | 344.08411 | 62.02817 | 0.76+/-0.1 | - | 12.0 | - | - | - | 29.1 | 66.8 | 89.7 | 69.1 | 0.0 | flat | - |

| Source ID | RA (2000) | Decl. (2000) | 2-10 keV cts ks$^{-1}$ | 0.4-2 keV cts ks$^{-1}$ | 2008 $L_X$ $10^{30}$ erg s$^{-1}$ | 2003 $L_X$ $10^{30}$ erg s$^{-1}$ | USNO B1 | 2MASS | IRAC1 mJy | IRAC2 mJy | IRAC3 mJy | IRAC4 mJy | Index | Class | Notes |
|---|---|---|---|---|---|---|---|---|---|---|---|---|---|---|---|
| s2 | 344.08440 | 62.03883 | - | 0.14+/-0.04 | 0.13 | 0.07 | 1520-0380898 | 22562025+6202196 | - | - | - | - | - | - | XMMPT J225620.4+620221, HL 28 |
| h4 | 344.08528 | 62.03301 | 0.11+/-0.04 | - | 0.15 | - | - | - | - | - | 191.7 | 99.5 | -3.0 | III | - |
| h3 | 344.08729 | 62.03445 | 0.53+/-0.08 | - | 0.75 | - | - | - | 95.8 | 270.9 | 326 | 249.2 | 0.1 | flat | - |
| h2 | 344.09248 | 62.03345 | 0.98+/-0.11 | 0.79+/-0.1 | 2.1 | 3.79 | - | - | - | 140.5 | - | - | - | - | XMMPT J225622.0+620204 |
| h1 | 344.09579 | 62.02802 | 0.09+/-0.04 | - | 0.13 | - | - | - | 5.9 | 7.9 | 7 | 4.3 | -1.5 | II | - |

[a]Offset from X-ray source by 1-1.2”, [b]Isophotal flux, [c]Flux estimate, [d]composite source, [e]0.4-10 keV

Table 2. Blankfield X-ray Sources

| Source ID | RA (2000) | Decl. (2000) | 2-10 keV cts ks$^{-1}$ | 0.4-2 keV cts ks$^{-1}$ | 2008 $L_X$ $10^{30}$ ergs s$^{-1}$ (0.4-10 keV) |
|---|---|---|---|---|---|
| s27 | 343.96585 | 62.03356 | - | 0.05+/-0.03 | 0.05 |
| s24 | 343.98888 | 62.04025 | - | 0.06+/-0.03 | 0.06 |
| s22 | 344.00004 | 62.07111 | - | 0.06+/-0.03 | 0.06 |
| h39 | 344.00081 | 62.05986 | 0.08+/-0.03 | - | 0.11 |
| s30 | 344.00281 | 62.02178 | - | 0.06+/-0.03 | 0.06 |
| h30 | 344.03224 | 62.03094 | 0.08+/-0.03 | - | 0.12 |
| s13 | 344.03597 | 62.01982 | - | 0.04+/-0.02 | 0.04 |
| h26 | 344.05023 | 62.04143 | 0.14+/-0.05 | - | 0.20 |
| h21 | 344.05505 | 62.03532 | 0.16+/-0.05 | - | 0.22 |
| h17 | 344.05880 | 62.02689 | 0.26+/-0.06 | - | 0.36 |
| s28 | 344.06557 | 62.03306 | - | 0.07+/-0.03 | 0.07 |
| h14 | 344.07251 | 62.03297 | 0.41+/-0.08 | - | 0.58 |
| h8 | 344.08042 | 62.02630 | 0.18+/-0.05 | - | 0.25 |

**Table 3: Spectral Parameters of Selected X-ray Sources in the Cep A Field**

| Source ID | Class | Net cts | kT (keV) | $N_H$ ($10^{23}$ cm$^{-2}$) | $I_{0.4\text{-}2\ keV}$ ($10^{-14}$ erg cm$^{-2}$ s$^{-1}$) | $I_{2\text{-}10\ keV}$ ($10^{-14}$ erg cm$^{-2}$ s$^{-1}$) | $L_X$ ($10^{30}$ erg s$^{-1}$) (0.4-10 keV) |
|---|---|---|---|---|---|---|---|
| HW 9 = h9 | - | 61 | > 0.5 | 0.5-5 | 0 (30) | 1.7 (9.9) | 1.0 (24) |
| HW 3c = h11 | - | 62 | > 5 | 6-14 | 0 (24) | 6.4 (59) | 3.8 (49) |
| h10 | - | 51 | >3 | 1.5-5.5 | 0 (3.0) | 2.9 (7.6) | 1.7 (6.2) |
| h19 | 0-I | 14 | ~10 | ~0.4 | 0 (-) | 0.6 (-) | ~0.26 (0.48) |
| h12 | 0-I | 20 | ~3.5 | ~0.5 | 0 (-) | 0.5 (-) | ~0.30 (0.88) |
| h7 | 0-I | 55 | >2 | 2-16 | 0 (7.4) | 4.3 (14) | 2.5 (13) |
| HL 8 = h38 | II | 114 5 | 2.3 ±0.4 | 0.079 (-0.014,+0.021) | 4 (16) | 9.5 (11) | 7.9 (16) |
| HL 9 = h36 | II | 371 | 1.9 (-0.4,+0.7) | 0.13 ± 0.04 | 1 (7.2) | 3 (3.7) | 2.4 (6.4) |
| h32 | II | 296 | 2.6 (-0.7,+1.4) | 0.35 (-0.08,+0.10) | 0.34 (11) | 6.2 (9.1) | 3.8 (12) |
| h20 | III | 177 | 1.8 (-0.5,+0.8) | 0.42 (-0.10,+0.16) | 1.8 (11) | 2.8 (5) | 1.8 (9.4) |
| h34 | III | 49 | 1-5 | 0.06-0.27 | 0.15 (1.5) | 0.41 (0.53) | 0.33 (1.2) |
| h6 | flat | 65 | 0.5-2.4 | 1.6 (-0.9,+1.5) | 0 (8.6) | 1.4 (12) | 0.82 (12) |
| h37 | - | 360 | 9 (-5,+40) | 0.30 ± 0.06 | 0.28 (5.3) | 9.3 (12) | 5.6 (10) |
| s23 | - | 42 | 0.1-3 | <0.13 | 0.21 (0.21) | 0.04 (0.04) | 0.15 (0.15) |
| HH 168 | HH object | 493 | 0.56 (+0.07,-0.05) | 0.057 (+0.013,-0.018) | 0.86 (4.9) | 0.07 (0.08) | 0.54 (3.0) |

**Table 4: Combined X-ray Spectra**

| ID | No. Srcs | Net cts | kT (keV) | $N_H$ ($10^{23}$ cm$^{-2}$) | *Total* $I_{0.4\text{-}2\ keV}$ ($10^{-14}$ erg cm$^{-2}$ s$^{-1}$) | *Total* $I_{2\text{-}10\ keV}$ ($10^{-14}$ erg cm$^{-2}$ s$^{-1}$) | *Total* $L_X$ ($10^{30}$ erg s$^{-1}$) (0.4-10 keV) |
|---|---|---|---|---|---|---|---|
| Power Sources | 3 | 201 | 7.5 (-3.7,+7.5) | 2.4 (-0.6,+1.1) | 0 (12) | 9.3 (25) | 5.5 (22) |
| Class 0-I | 3 | 89 | >3 | 1.1 (-0.6,+1.1) | 0 (2.7) | 3.3 (5.9) | 1.9 (5.0) |
| Class II | 10 | 406 | 3.5 (-1.2,+17) | 0.18 (-0.09,+0.07) | 0. 68 (6.4) | 5.9 (7.0) | 3.9 (7.9) |
| Class III hard | 7 | 317 | 2.3 (-0.6,+1.0) | 0.17 (-0.04,+0.05) | 0.69 (6.7) | 3.6 (4.5) | 2.5 (6.6) |
| Blankfield hard | 7 | 228 | >7 | 0.15 (-0.06,+0.13) | 0.16 (0.99) | 3.0 (3.3) | 1.9 (2.5) |
| Blankfield soft | 6 | 67 | >2 | <0.04 | 0.17 (0.17) | 0.30 (0.30) | 0.28 (0.28) |

## FIGURE CAPTIONS

1. The *Chandra/ACIS* X-ray image from CCD I3 in the 0.4-10 keV energy band. The green 1"-radius circles mark the positions of X-ray sources detected with the wavelet analysis that have counterparts at other wavelengths (see Table 1). The cyan 1"-radius circles mark the positions of "blankfield" X-ray sources (Table 2). The box shows the location of the putative power source(s) and HWX (PT05).

2. Spitzer/IRAC images of the Cepheus A East region in four bands—A: 3.6 μm; B: 4.5 μm; C: 5.8 μm; and D: 8.0 μm. The green 6" by 6" squares mark the sources detected in the hard band (at least) with counterparts at other wavelengths; green 3"-radius circles are sources detected in the soft band only with counterparts at other wavelengths; cyan 3"-radius circles are "blankfield" sources. For the labels of sources in the power source region, refer to Fig.3.

3. Same as Fig. 2 but expanded view of the power source region.

4. a) The *Chandra/ACIS* X-ray image of Cepheus A East exciting region in the 2-8 keV energy band. The green 1"-radius circles mark the positions of X-ray sources detected with the wavelet analysis that have counterparts at other wavelengths (see Table 1). The cyan 1"-radius circles mark the positions of "blankfield" X-ray sources (Table 2). The 0.5"-radius magenta circles mark the positions of the compact radio "HW" sources (Garay et al. 1996).

   b). A blowup of the central region of Fig. 4a. The 0.2"-green circles mark the locations of compact radio "HW" sources (Torrelles et al. 1998), the magenta circle is HW8 (Garay et al. 1996), and the smaller blue circles mark the positions of SMA sources (Brogan et al. 2007).

5. X-ray data and model spectrum of the total emission from the Cepheus A East core region.

6. *Chandra/ACIS* light curves for selected X-ray sources in Cepheus A: a) HW9 (h9), b) HW3c (h11), c) h10, d) HL 8 (h38), e) HL 9 (h36), and f) h37.

7. X-ray data and model spectrum of the Class II object, HL 8.

8. *Chandra/ACIS* image of HH 168 with X-rays < 2 keV (red) and >2 keV (cyan)

9. The *Chandra/ACIS* X-ray contours (green) superposed upon the *HST* WFPC2 image of Cepheus A taken with the Hα filter (Hartigan et al. 2000). The circles are compact radio sources: blue circles from Hughes & Moriarty-Shieven (1990), and magenta circles are from Garay et al. (1996). The white squares show the X-ray localizations from PT05.

10. The *Chandra/ACIS* X-ray data and best-fit model (black), the XMM-Newton data and the same best-fit model (MOS 1+2; red, PN; green).

11. Radio luminosity vs. unabsorbed X-ray luminosity of HH objects (points) and best fit linear relation (line).

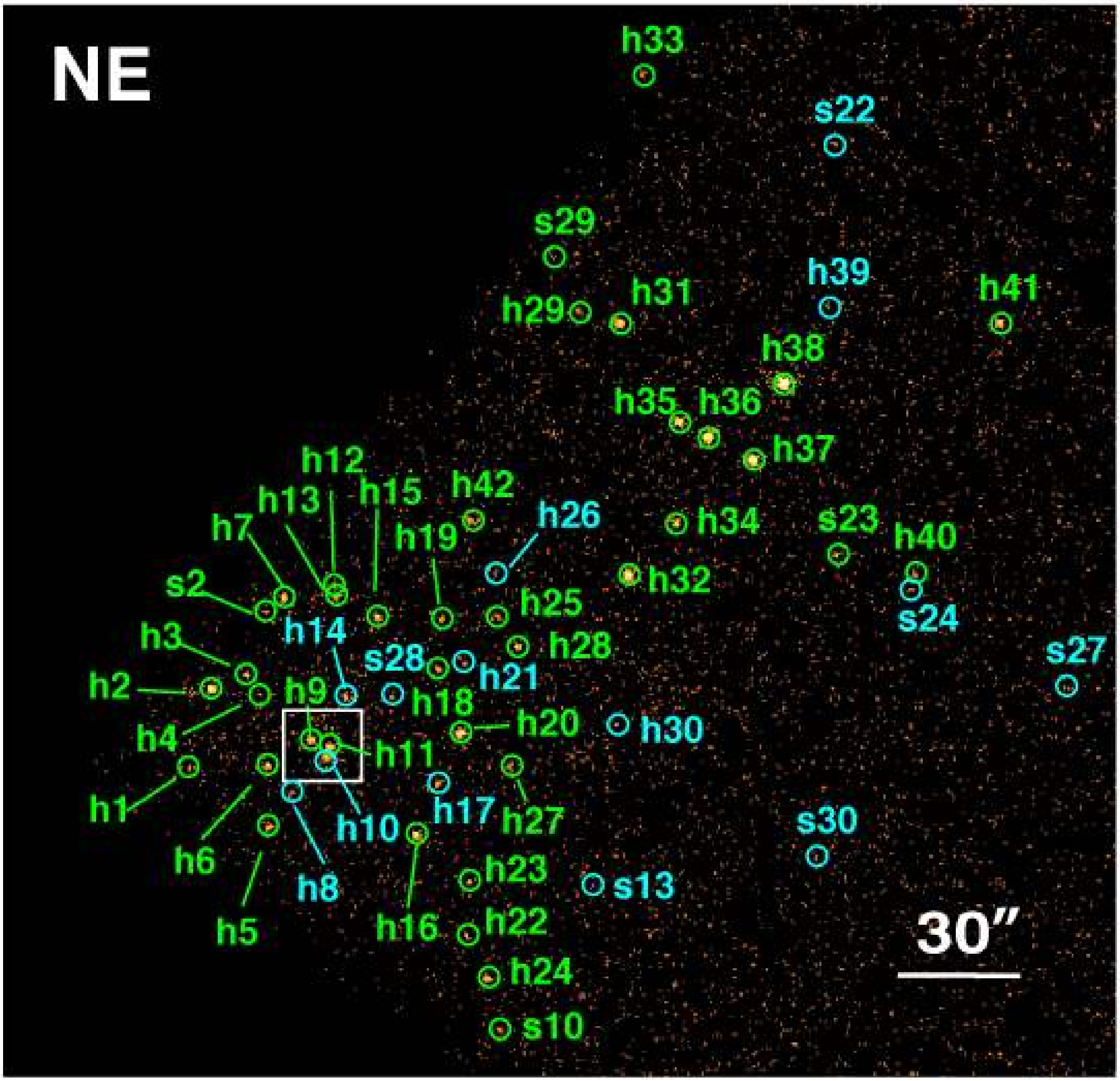

**Figure 1**

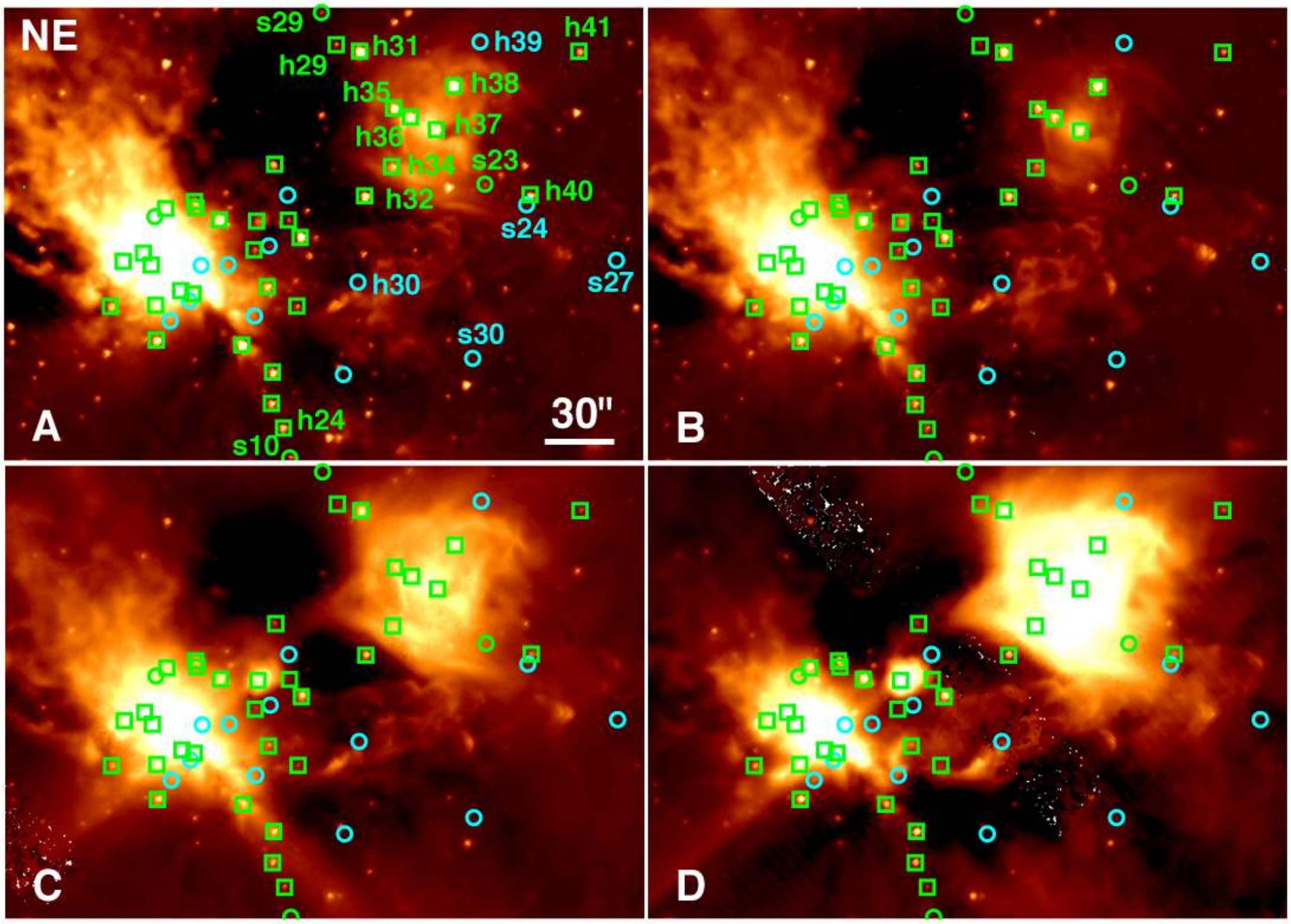

**Figure 2**

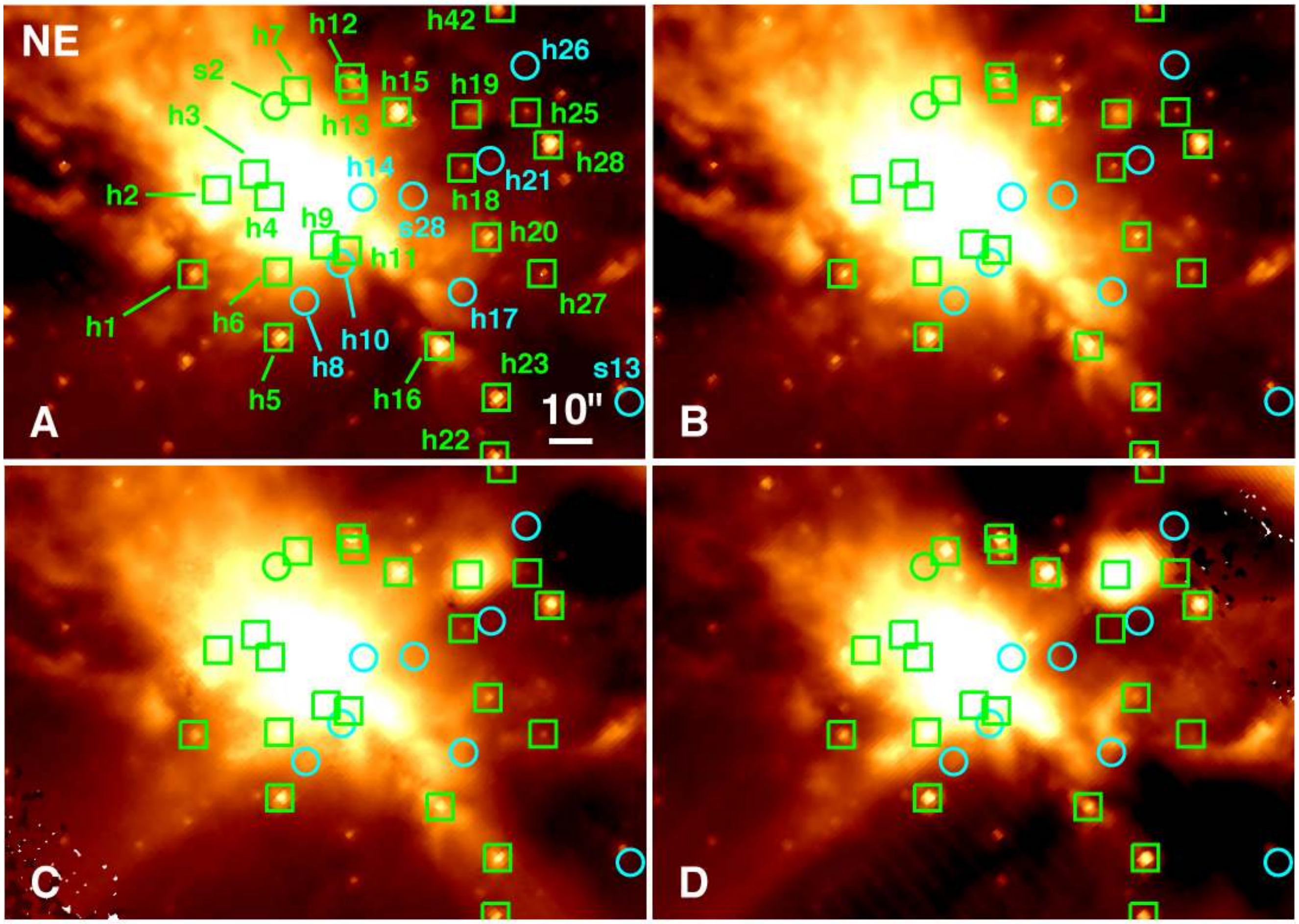

**Figure 3**

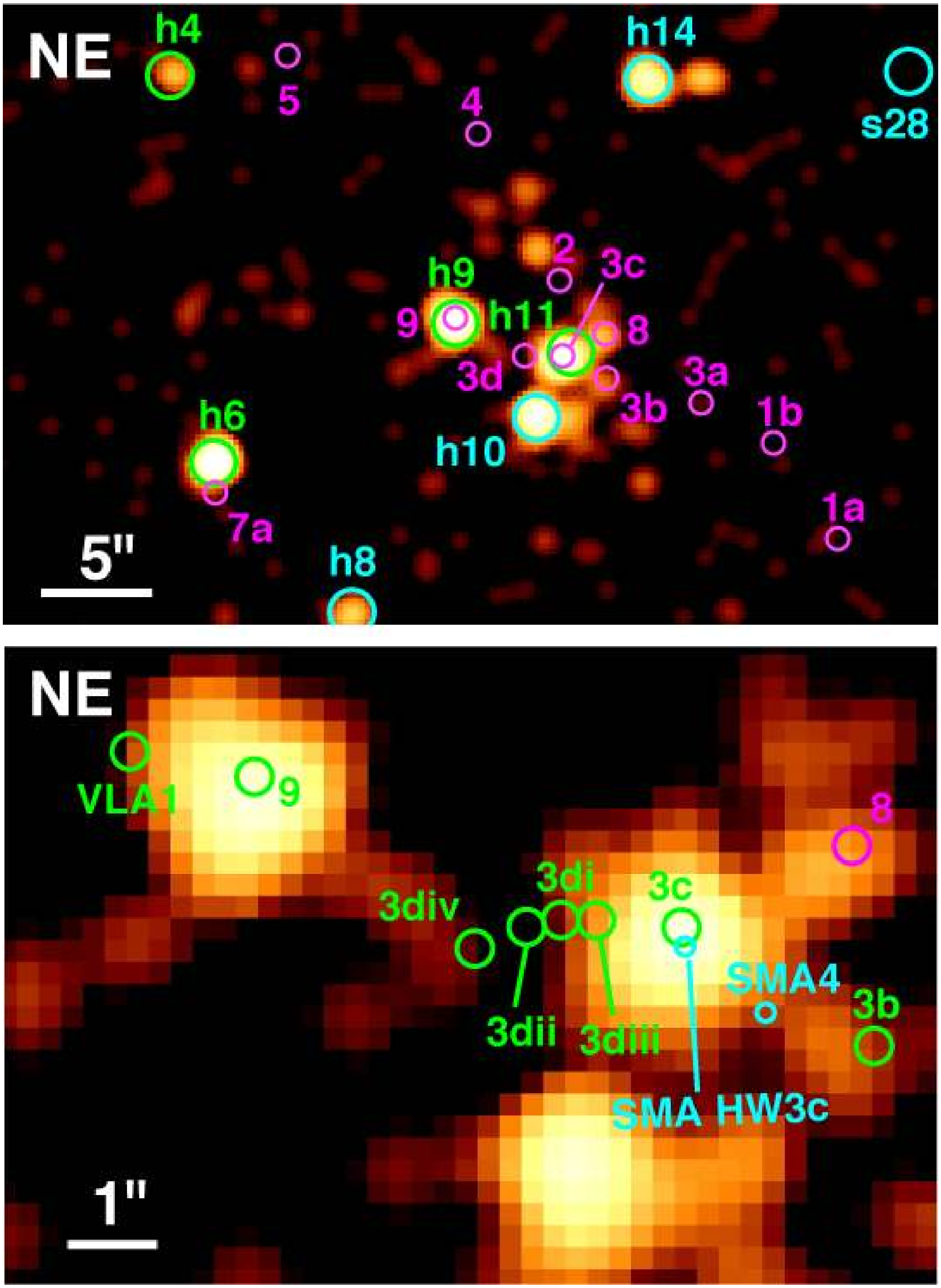

**Figure 4a (above), b(below)**

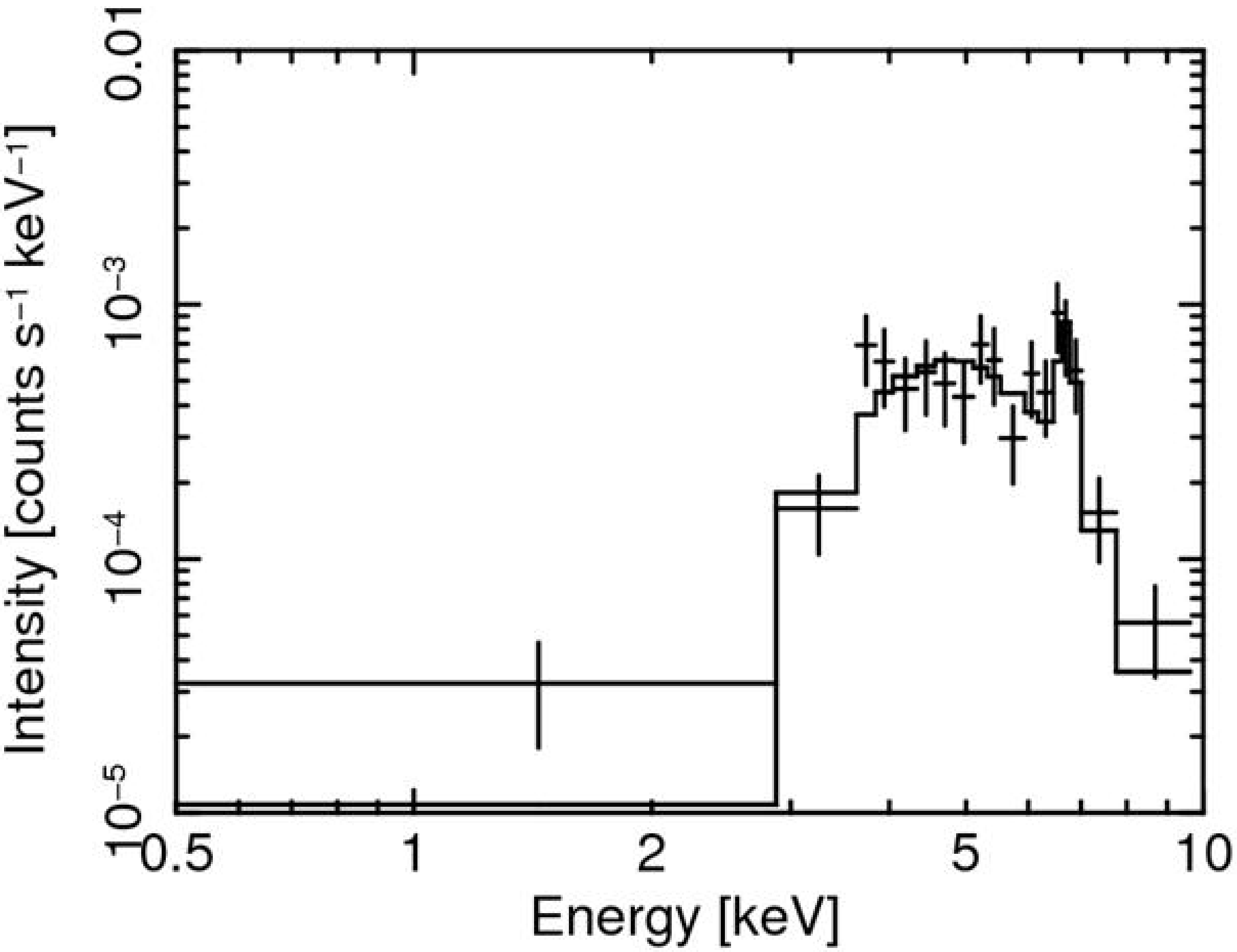

**Figure 5**

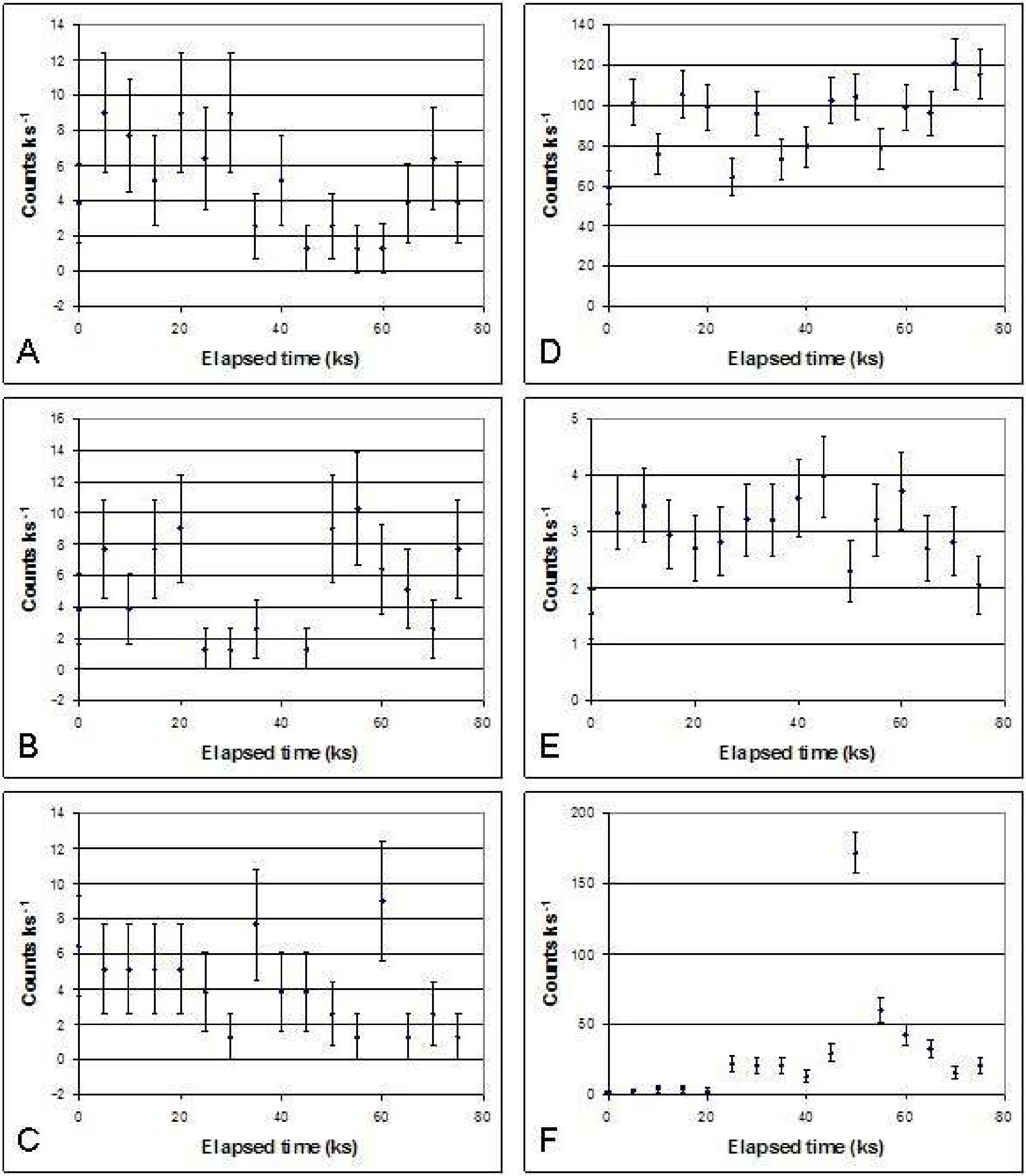

**Figure 6**

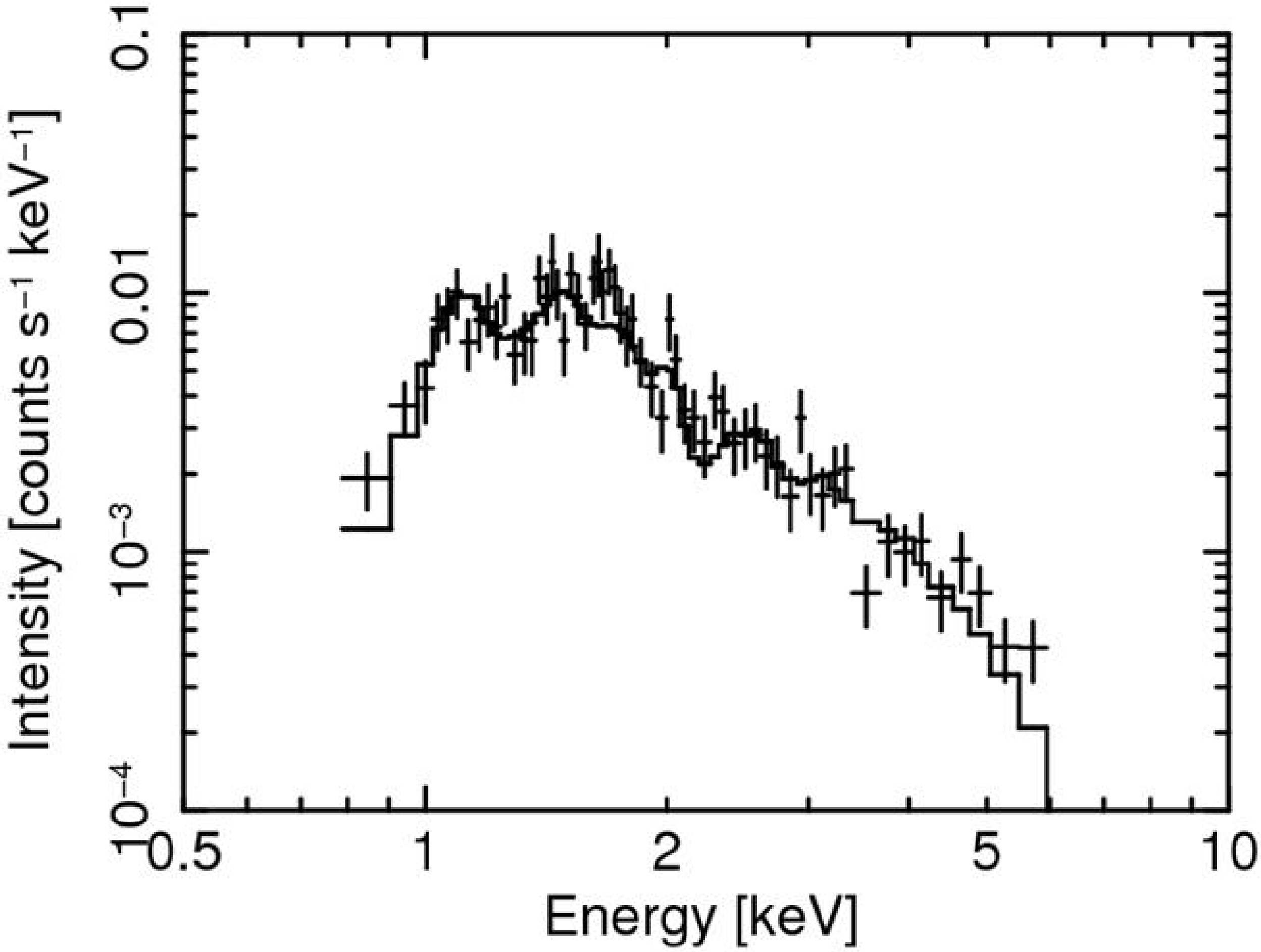

**Figure 7**

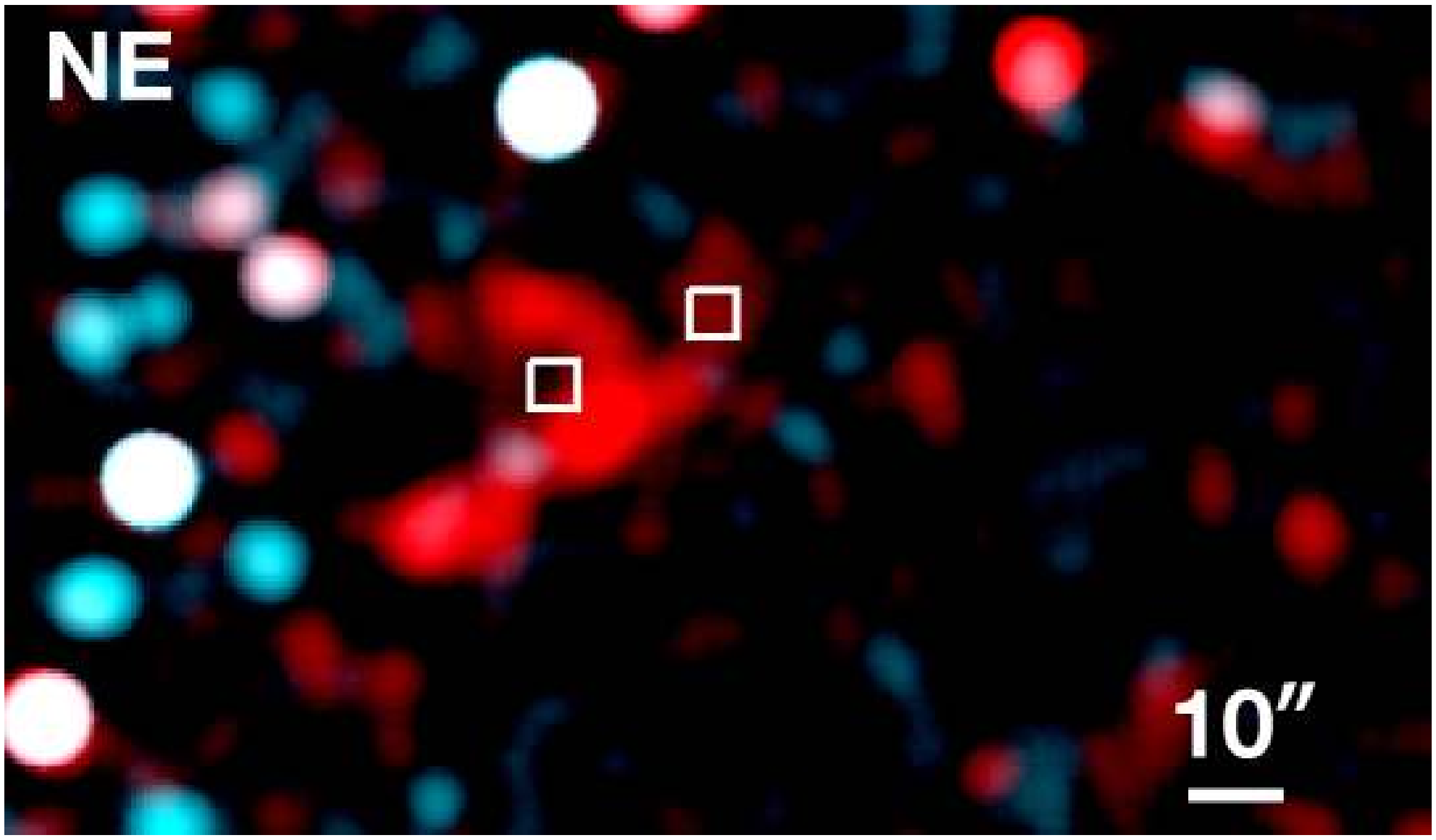

**Figure 8**

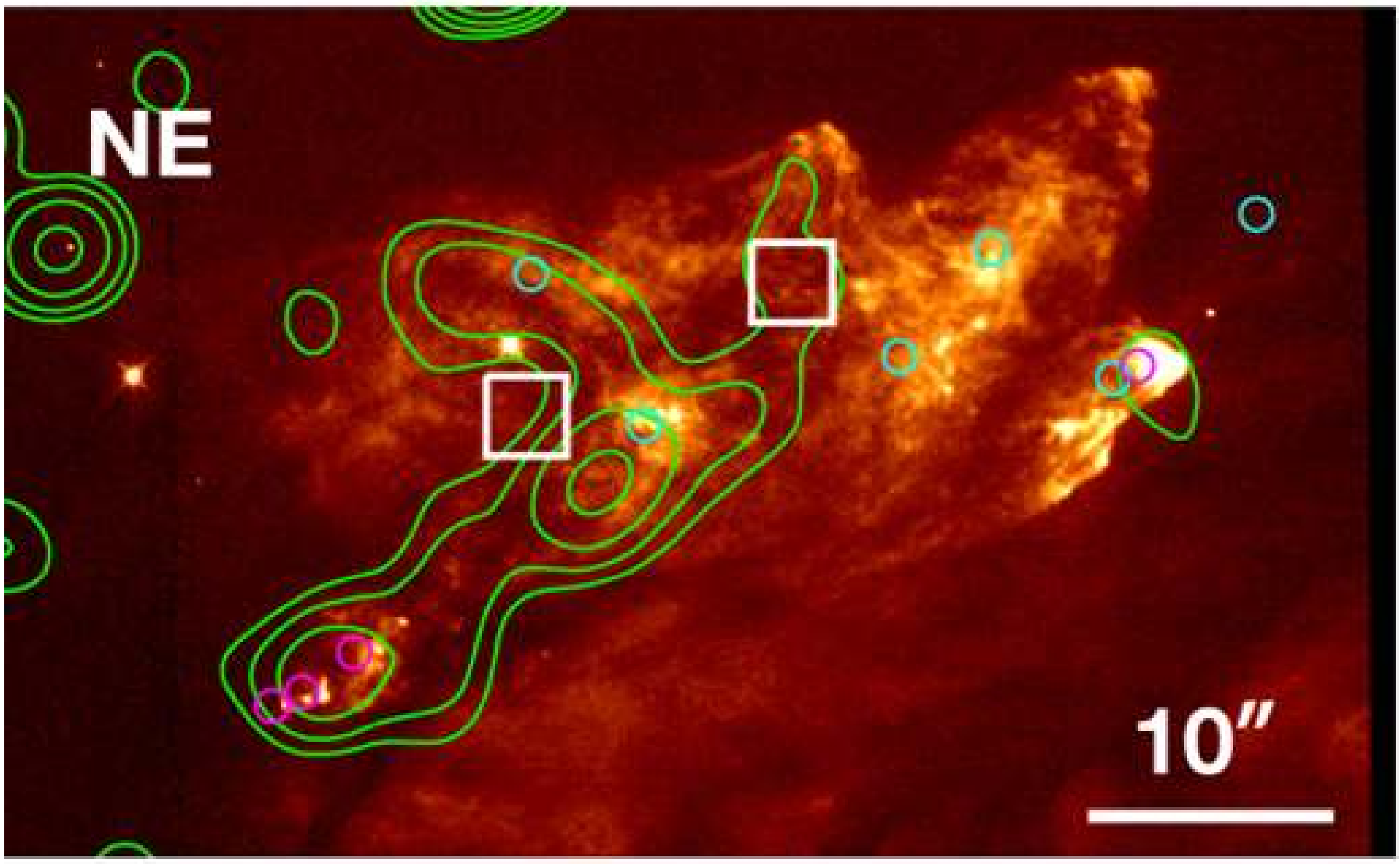

**Figure 9**

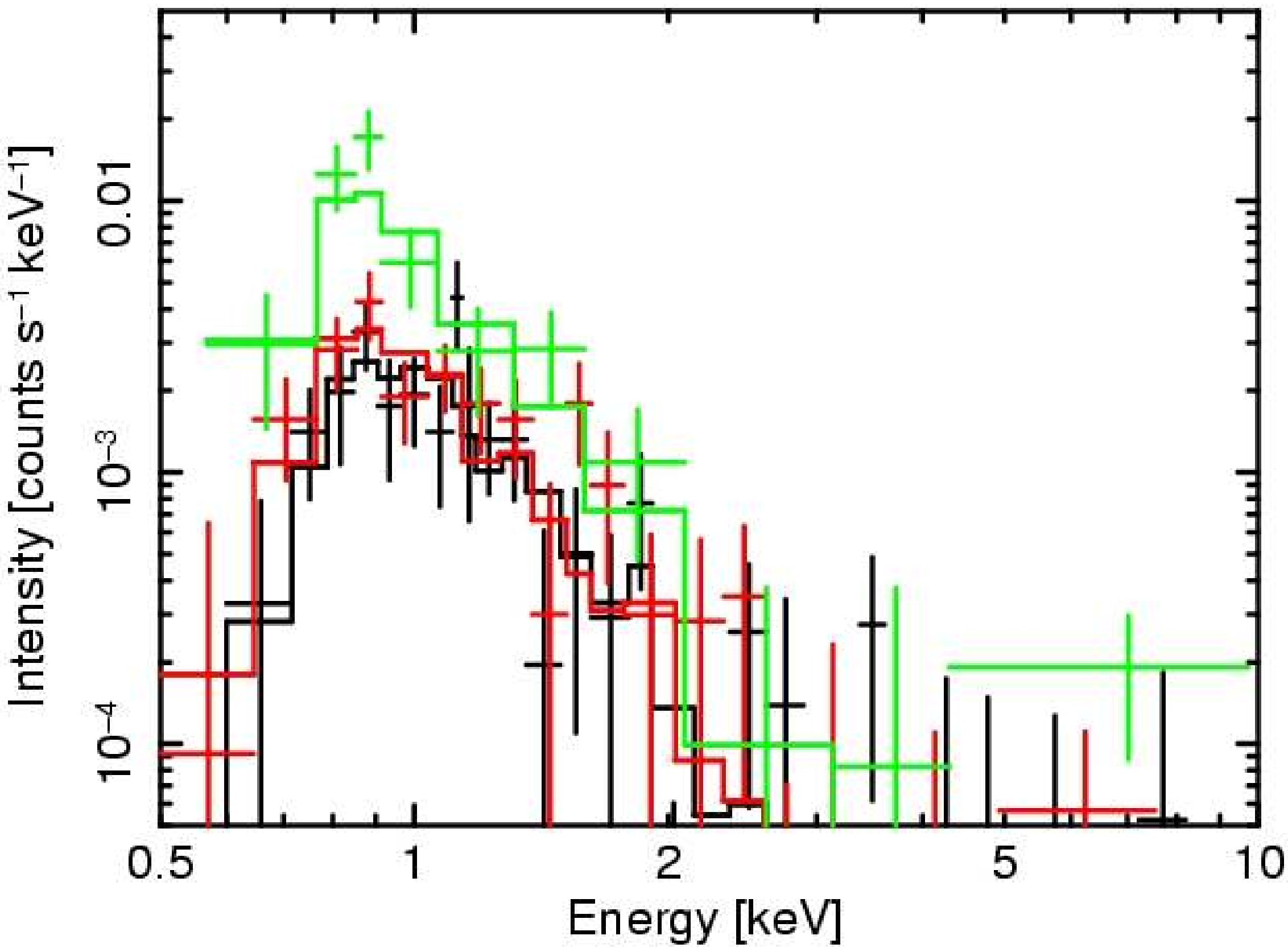

**Figure 10**

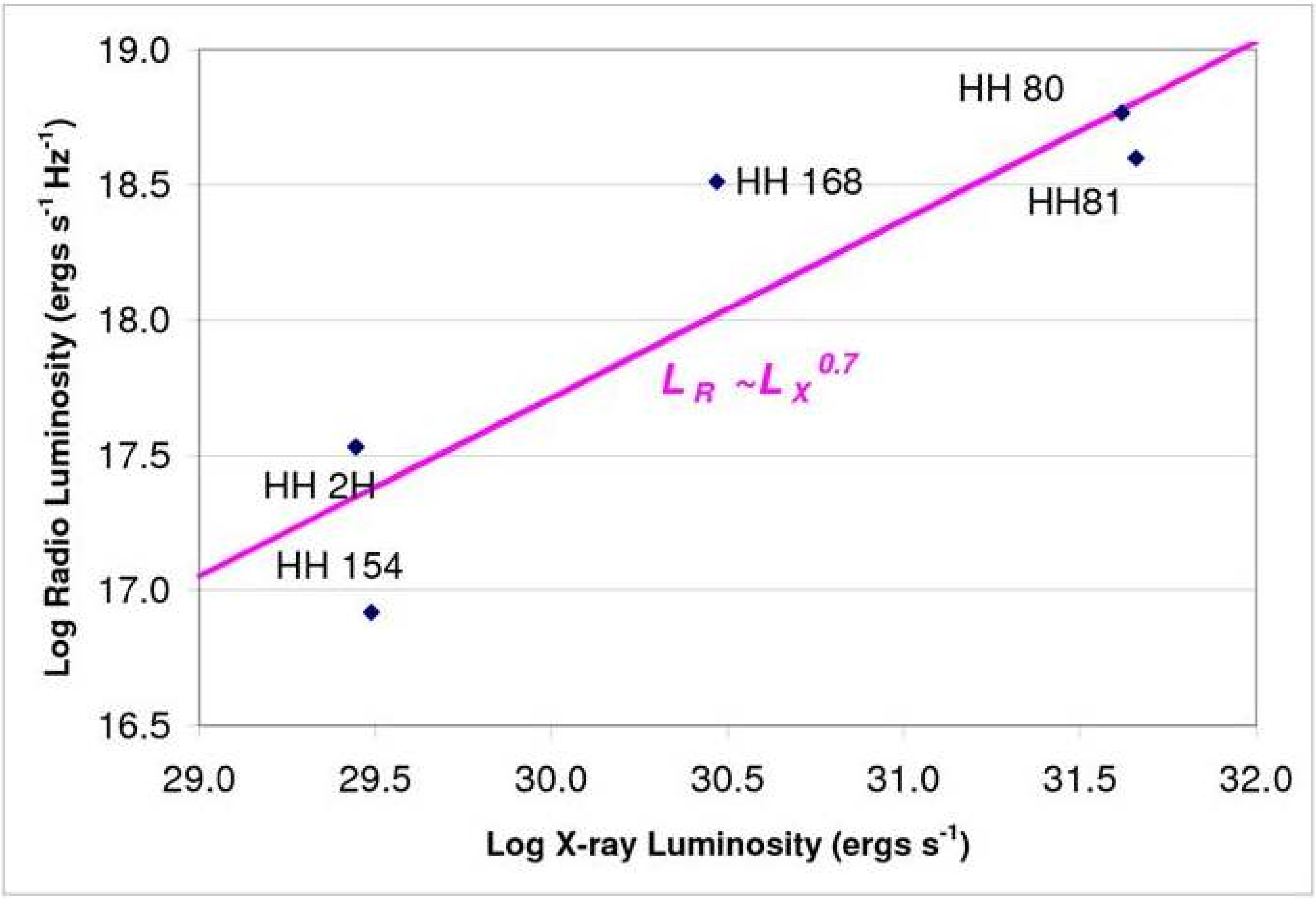

**Figure 11**